\documentstyle[12pt,a4]{article}

\begin{document}
\title{Study by the Methods of Physics of some Medieval Coins}
\author{Agata Olariu\\
{\em National Institute for Physics and Nuclear Engineering}\\
{\em P.O. Box MG-6 76900 Bucharest Magurele, Romania}}

\maketitle
\section{Introduction}

In this paper we have analyzed a number of medieval coins from the region of 
Sibiu, Romania. The compositional analyses have been made in conjunction with 
the 
dicovery in a pit, at the Old Town from Sibiu of 2 ovens for smelting of copper
, with some metal
pieces on the surface of them$^{1}$. The archaeologists believed that in this
place was discovered the old mint from Sibiu.  To verify this hypothesis 
some coins from the collection of the Brukenthal Museum - Sibiu, have been 
analyzed to be compared later to the composition of the
metal pieces from the 2 ovens. For the investigation of the composition of the
coins we have applied 3 methods: , $\gamma$
transmission$^{2}$, X-ray fluorescence (XRF) and neutron activation analysis
(NAA).
In the Table 1 is shown the list of analyzed coins.
\begin{table}[h]
\newlabel{}
\caption{{\bf Table 1.} List of ancient analyzed coins, Brukenthal Museum, 
Sibiu}\\

\begin{tabular}{lllc}
\hline
Coin  & Reg. no. & Characteristics & Provenance\\
\hline
\hline\\
Matei Corvin denarius, Fig. 1 &T.1285/4641 & common metal with silver& Sibiu\\
Transylvania denarius, Fig. 2  & T.1285/2455 & common  metal         & Sibiu\\
XV cent. denarius           &  fragment    & from arch. sett.     & Saliste\\
Piece Leopold I, Fig. 3 & T.1285/3762 & common metal with silver     & Kremnig\\
 15 Kr, 1704            &              &                       &       \\
\hline
\end{tabular}
\end{table}
\\
\newpage
\section{The methods of analysis}
{\it $\gamma$ Transmission}\\
was carried out with a source of $^{241}$Am which gives a intense 
$\gamma$ line of 
 59.54 keV, at a $\gamma$ spectrometer formed of a Ge(Li) detector, 135cm$^3$, 
and a PC with MCA interface. Considering that the coins have a binary 
composition (Ag + Cu) we have made a set of standards:  sandwiches of 
small discs of pure copper and pure silver; some known Romanian coins of known
composition, Carol I, 50 bani (Ag 83.5\%, Cu 16.5\%), small coins of 
7 bani with Cu 100\% and 4 bani with Ag 100\%.
The sandwiches of discs can carry out the imitation of binary coins of Cu 
and Ag with
known concentrations. This method can give information of the bulk of the coin.
To have a complete information of the composition and the homogeneity of the
coin, we have performed different measurements of the coin put in
different geometry positions in respect to the endcap of the detector 
of $\gamma$ rays. 
If between the source of $^{241}$Am and the $\gamma$ detector it is
interposed a binary coin, of 2 materials, with 2 different densities
and with 2 different attenuation coefficients: 
  $\mu_1$ 'si $\mu_2$, then the photopeakul of 59.54 keV is attenuated 
on the equation:
\begin{center}
{\large       I$_x$=I$_0$e$^{-(\mu x_1+\mu x_2)}$}
\end{center}
We have made a calibration (Fig. 4) with the aid of above mentioned standards 
and we have determined the function:
\begin{center}
{\large          Transmission$_{\gamma}$=f(Concentrantion$_{Ag}$ or $_{Cu}$)}
\end{center}
or explicitly:
\large
\begin{center}
$\frac{1}{M/S}$ln$\frac{I_0}{I_x}$=C$_2$($\mu_2\prime$-$\mu_1\prime$)+$\mu_1\prime$
\end{center}
\normalsize
where M, is the mass of the standard, coin\\
S, surface of the standard, coin\\
I$_0$, aria of photopeak of 59.54 keV, without absorber\\
I$_x$, aria of photopeak of  59.54 keV, with attenuator- standard, coin\\
C$_2$, concentration of Ag\\
$\mu_1\prime$=$\mu_1\cdot\rho_1$\\
$\mu_2\prime$=$\mu_2\cdot\rho_2$\\
with $\rho$, density.\\

{\it XRF}\\
The same coins have been analyzed by XRF using a triple, ring-shaped  
excitation source of $^{238}$Pu,
and for the detection of the X-ray we have used a spectrometer chain formed 
by a GeHP detector, Model GL011OP-7905SL-15, Horizontal Integral, with 
sensible
volume of 100x10 mm, window of 0.075 mm and a resolution of 165 eV at 5.9 keV
($^{55}$Fe) . The acquisition of X rays of fluorescence have been made on-line 
on a PC with a MCA interface.
We have prepared flattened standards for XRF, of known composition.
The results of the XRF analysis of the ancient coins from Brukenthal Museum 
are given in the Table 3. We considered the statistical errors which are 
$<$3\% . We also considered  that the coins are binary
and we have made corrections to have C$_{Ag}$ + C$_{Cu}$=100\%. 
XRF method gives information on the surface of the coins and from the 
comparison of the Table 2 and Table 3 we can observe that the concentration of
silver at the surface of the coin has larger values, effect known in the
literature $^{3}$. \\

{\it NAA}\\
We applied the NAA method for 2 coins, which permitted to take samples, without
damaging the coins. 
The samples of $\approx$ 1-2 mg and a standard of flux have been irradiated at
the rabbit system of the VVR-S reactor from NIPNE Magurele, Bucharest,
for a period of 75 min. 
The measurement of the radioactivity induced in the samples have been performed
after 1 d of decay, at the  $\gamma$ spectrometer described above, at the 
$\gamma$ transmission method. We have determined the following elements: Ag,
Cu, Au, As and Sb.
The results of the NAA on the 2 coins analyzed are given in the Table 4. the
errors of measurement for the elements determined are $<$5\%.

\section{Results and discussions}

In the Table 2 are shown the results of the $\gamma$ transmission analysis 
for the ancient coins from Brukenthal Museum, Sibiu. For the same coin are 
shown different concentrations,
obtained from various measurements, corresponding to different points 
on the surface of the coin.
\begin{table}[h]
\newlabel{}
\caption{{\bf Table 2.} Concentrations of Ag and Cu in \%,  coins from\\
Brukenthal Museum Sibiu, by $\gamma$ Transmission}\\
\begin{tabular}{llcc}
\hline
Coin  & Reg. no. & Ag, \% & Cu, \%\\
\hline
\hline
Matei Corvin denarius  &T.1285/4641  & 57.5 & 42.5 \\
Matei Corvin denarius  &             & 68.0 & 32\\
Matei Corvin denarius  &             & 70.5 & 29.5\\
\hline
Transylvania  denarius & T.1285/2455 & 35.75 & 64.25\\
Transylvania  denarius &             & 37.3 & 62.7 \\
\hline
XV cent. denarius     & Saliste& 100 &  0\\
XV cent. denarius     & Saliste& 94.5 & 5.5 \\
\hline
Piece Leopold I    & T.1285/3762  & 57   & 43  \\
Piece Leopold I    &              &  54  & 46  \\
Piece Leopold I    &              & 51.5 & 48.5 \\
Piece Leopold I    &              & 57   & 43  \\
Piece Leopold I    &              & 60   & 40  \\
Piece Leopold I    &              & 52.5 & 47.5 \\
\hline
\end{tabular}
\end{table}

\begin{table}
\newlabel{}
\caption{{\bf Table 3} Concentration of Ag and Cu, \%, for analyzed coins, 
by XRF}\\

\begin{tabular}{lcc}
\hline
Coin  & Ag, \% & Cu, \%\\
\hline
\hline\\
Matei Corvin denarius  	   & 88.2   & 11.8      \\
Transylvania denarius      & 51.0   & 49.0      \\
XV cent. denarius, Saliste & 96.1   &  3.9      \\
Piece of 15 Kr. Leopold I  & 77.2   & 22.8      \\
                           &        &           \\
\hline
\end{tabular}
\end{table}

\begin{table}
\newlabel{}
\caption{{\bf Table 4.} Concentrations for 2 coins from Brukenthal Museum, 
Sibiu, by NAA}\\

\begin{tabular}{lccccc}
\hline
Coin & Cu & Ag & Au & As & Sb \\
     & \% & \% & ppm & ppm& ppm\\  
\hline
\hline\\
Transylvania denarius       & 68.8 & 27.3 & 1360 & 2340 & 2050      \\
XV cent., Saliste denarius  & 3.5  & 96.5 & 59   & n.d. & 3180  \\
                            &       &     &      &      &       \\
\hline
\end{tabular}
\end{table}
The $\gamma$ transmission as well as the NAA give information about the bulk
content of the coins. XRF gives information about the surface composition of
the coins. There are different results for the Cu and Ag concentrations for 
Transylvanian denarius and Saliste coin and this situation could be explain by
the fact the coins are inhomogeneous. From this point of view the $\gamma$ 
transmission can characterize better the coin. 
However NAA gives supplementary information on the trace elements from the
composition of coins. The analyses have changed completely the image on the
Transylvanian denarius, considered as "common metal". The analyses show that it
contains about 35\% of Ag.\\

{\bf Acknowledgement}\\
I thank the archaeologist Petre Besliu-Munteanu for 
collaboration in doing this work.
\newpage
{\bf \large References}

\noindent
1. Agata Olariu and P. Besliu-Munteanu, Characterization of fourteenth-century 
bell-casting pit in Old Town in Sibiu, Romania, Bulletin of Historic Monuments, 
No. 1-2 (1997) and Los Alamos e-print Archive nucl-ex, paper 9908006\\
2. V. Cojocaru, Gamma transmission used in binary coins analysis, 
Annual Report, Department of Heavy Ion Physics, Bucharest, 1986, p. 103\\
3. J. Hughes, Surface and Interior Analysis of Antiquities, Nucl. Instr. 
Methods in Phys. Res., B14 (1986) p. 16 \\
\end{document}